\documentclass[ALICE,manyauthors]{cernphprep}

\def\pt {\mbox{$p_{\rm T}$}\xspace}    
\def\pbpb {Pb--Pb\xspace}
\def\ppb {p--Pb\xspace}

\def\pp {pp\xspace}

\def\upsi {\mbox{$\Upsilon$}\xspace}
\def\upsis {\mbox{$\Upsilon$(1S)}\xspace}
\def\upsiss {\mbox{$\Upsilon$(2S)}\xspace}
\def\upsisss {\mbox{$\Upsilon$(3S)}\xspace}
\def\raa {\mbox{$R_{\rm{AA}}$}}
\def\taa {\mbox{$\langle T_{\rm{AA}} \rangle$}\xspace}

\newcommand{\snn}  {\ensuremath{\sqrt{s_{_{\rm NN}}}}}   
\def\npart {\mbox{$N_{\rm{part}}$}}

\usepackage{datetime}
\usepackage{amsmath}
\usepackage{xcolor}
\usepackage[comma,square,numbers,sort&compress]{natbib}
\usepackage{hyperref}
\usepackage{lineno}
\usepackage{multirow}
\usepackage{gensymb}

\begin{document}%

\begin{titlepage}
\PHyear{2018}
\PHnumber{114}       
\PHdate{12 May}  

%

\title{$\Upsilon$ suppression at forward rapidity in Pb--Pb collisions \\ at  $\mathbf{\sqrt{\textit{s}_{_{\rm NN}}} = 5.02}$~TeV}
\ShortTitle{$\Upsilon$ suppression at forward rapidity in Pb--Pb collisions at  $\sqrt{s_{_{\rm NN}}} = 5.02$~TeV}   

\Collaboration{ALICE Collaboration\thanks{See Appendix~\ref{app:collab} for the list of collaboration members}}
\ShortAuthor{ALICE Collaboration} 

\begin{abstract}
Inclusive \upsis and \upsiss production have been measured in \pbpb collisions at the centre-of-mass energy per nucleon-nucleon pair $\sqrt{s_{_{\rm NN}}}=5.02$ TeV, using the ALICE detector at the CERN LHC.
The \upsi mesons are reconstructed in the centre-of-mass rapidity interval $2.5<y<4$ and in the transverse-momentum range $p_{\rm T}<15$ GeV/$c$, via their decays to muon pairs.
In this Letter, we present results on the inclusive \upsis nuclear modification factor $R_{\rm AA}$  as a function of collision centrality, transverse momentum and rapidity. 
The \upsis and \upsiss $R_{\rm AA}$, integrated over the centrality range 0--90\%, are  $0.37 \pm 0.02 {\rm{(stat)}}\pm 0.03 {\rm{(syst)}}$ and $0.10 \pm 0.04 {\rm{(stat)}}\pm 0.02 {\rm{(syst)}}$, respectively, leading to a ratio $R_{\rm{AA}}^{\Upsilon(\rm2S)}/R_{\rm{AA}}^{\Upsilon(\rm1S)}$ of $0.28\pm0.12\text{(stat)}\pm0.06\text{(syst)}$.
The observed \upsis suppression increases with the centrality of the collision and no significant variation is observed as a function of transverse momentum and rapidity.

\end{abstract}
\end{titlepage}
\setcounter{page}{2}

%
%

\section{Introduction}
A detailed study of the properties of the Quark-Gluon Plasma (QGP)~\cite{Shuryak:1978ij} is the main goal of heavy-ion experiments at ultra-relativistic energies~\cite{Arsene:2004fa,Back:2004je,Adams:2005dq,Adcox:2004mh,Muller:2012zq}. 
Quarkonia, $i.e.$ bound states of charm or bottom quark-antiquark pairs, are sensitive probes of color deconfinement, due to the Quantum-Chromo Dynamics Debye screening mechanism  \cite{Matsui:1986dk,Brambilla:2010cs,Andronic:2015wma} leading to quarkonium suppression. 
Moreover, the various quarkonium states have different binding energies and therefore different dissociation temperatures in a QGP, leading to sequential suppression ~\cite{Matsui:1986dk,Digal:2001ue}. 
Theory estimates \cite{Krouppa:2015yoa} indicate that bottomonium formation may occur before QGP thermalization \cite{Mauricio:2007vz} because of the large bottom quark mass. 
In this situation, a quantitative description of the influence of the medium on the bound states becomes challenging.
While the dissociation temperatures vary significantly between different models \cite{Brambilla:2010cs,Andronic:2015wma}, it is commonly accepted that the widths of the spectral functions of the bottomonium states increase compared to the widths in vacuum, due to the high temperature of the surrounding medium~\cite{Burnier:2014ssa}. 
Finally, taking into account that feed-down processes from higher-mass resonances (around 40\% for the \upsis~and 30\% for the \upsiss \cite{Andronic:2015wma}) are not negligible, the evaluation of the medium temperature via bottomonium measurements remains a complex endeavour.

The first studies of quarkonium production in heavy-ion collisions were devoted to charmonium states, and a suppression of their yields was observed
at the SPS~\cite{abreu:in2p3-00002434,Alessandro:2004ap,Arnaldi:2007zz}, at RHIC~\cite{Adare:2011yf,Abelev:2009qaa} and at the LHC~\cite{Abelev:2012rv,Abelev:2013ila,Khachatryan:2016ypw,Adam:2016rdg}. 
The weaker J/$\psi$ suppression observed at LHC energies, where the centre-of-mass energy per nucleon-nucleon pair (\snn) is one order of magnitude larger than at RHIC, can be explained by means of a competitive (re)generation mechanism, which occurs during the deconfined phase and/or at the hadronization stage~\cite{BraunMunzinger:2000px,Thews:2000rj,Zhao:2011cv,Zhou:2014kka}. 
This production mechanism strongly depends on the (re)combination probability of deconfined quarks present in the medium and thus on the initial number of produced ${\rm c{\overline c}}$ pairs. 
The effect has been found to be more important at low $p_{\rm T}$ and in the most central collisions \cite{Adam:2016rdg,Abelev:2013ila,Adam:2015isa}. 

The high-energy collisions delivered by the LHC allow for a detailed study of bottomonium states. 
For bottomonium production, perturbative calculations of production rates in elementary nucleon-nucleon collisions are more reliable than for charmonium yields due to the higher mass of the bottom quark with respect to charm~\cite{Bodwin:1994jh}. 
Since the number of produced ${\rm b{\overline b}}$ pairs in central heavy-ion collisions amount to a few pairs per event at the LHC, the probability for (re)generation of bottomonia through (re)combination is much smaller than in the case of charmonia.

The \upsis nuclear modification factor $R_{\rm AA}$ is quantified as the ratio of the \upsis yield in nucleus--nucleus collisions to the production cross section measured in pp collisions scaled by the nuclear overlap function \taa. 
The latter is obtained via the Glauber model~\cite{Miller:2007ri,Loizides:2017ack}. 
A strong suppression of the \upsis state in \pbpb collisions has been observed at $\sqrt{s_{_{\rm NN}}}=2.76$ TeV by ALICE \cite{Abelev:2014nua} and CMS \cite{Chatrchyan:2012lxa,Khachatryan:2016xxp} in the rapidity ranges $2.5<y<4$ and $|y|<2.4$, respectively. 
The suppression increases with the centrality of the collision, reaching about 60\% and 80\% for the most central collisions at mid~\cite{Khachatryan:2016xxp} and forward rapidity~\cite{Abelev:2014nua}, respectively. 
Moreover, the \upsiss suppression reaches about 90\% and for \upsisss data are compatible with a complete suppression \cite{Khachatryan:2016xxp}. 
As a function of \pt the \upsis \raa, measured for $p_{\rm T}<20$ GeV/$c$ by CMS~\cite{Khachatryan:2016xxp}, is compatible with a constant value. 
When considering the $y$-dependence resulting from the comparison of ALICE and CMS results, there is an indication for a stronger suppression at forward $y$.
Transport models \cite{Zhou:2014kka,Du:2017qkv} as well as an anisotropic hydro-dynamical model \cite{Krouppa:2017jlg} fairly reproduce the experimental observations of CMS, while they tend to overestimate the $R_{\rm AA}$ values measured by ALICE. 

The bottomonium suppression due to the QGP should be disentangled from the suppression due to Cold Nuclear Matter (CNM) effects, such as the nuclear modification of the parton distribution functions due to shadowing \cite{Eskola:1998df,Eskola:2009uj}, as well as parton energy loss \cite{Arleo:2012rs}. 
These effects on the bottomonium production were studied in \ppb collisions by ALICE \cite{Abelev:2014oea} and LHCb \cite{Aaij:2014mza}, who reported for the \upsis a nuclear modification factor slightly lower than unity at forward rapidity and compatible with unity at backward rapidity, although with significant uncertainties. 
Recently, ATLAS results indicate a significant suppression of the \upsis for $p_{\rm T}$ $<$ 40 GeV/$c$ around mid-rapidity~\cite{Aaboud:2017cif}.  
Additional measurements at forward/backward rapidity with higher statistics, are needed to fully constrain the models and perform a meaningful extrapolation of CNM effects to \pbpb collisions.

In this Letter we present the first results on the \upsis and \upsiss $R_{\rm AA}$ measured by the ALICE Collaboration in \pbpb collisions at $\sqrt{s_{_{\rm NN}}}=5.02$ TeV. 
The pp reference cross sections used in the $R_{\rm AA}$ calculations have been determined by an interpolation procedure based on various ALICE \cite{Abelev:2014qha,Adam:2015rta} and LHCb \cite{Aaij:2014nwa,Aaij:2015awa} results at different energies. 
The nuclear modification factor for the \upsis is presented as a function of the centrality of the collision and also differentially in $p_{\rm T}$ and rapidity.
For the \upsiss, an $R_{\rm AA}$ value integrated over the centrality of the collision is quoted. 
Finally, the results are compared to theoretical calculations.

\section{Experimental apparatus and data sample}
\label{section:Apparatus}

An extensive description of the ALICE apparatus can be found in \cite{Aamodt:2008zz,Abelev:2014ffa}.
The analysis presented in this Letter is based on muons detected at forward rapidity ($2.5<y<4$)\footnote{In the ALICE reference frame, the muon spectrometer covers a negative $\eta$ range and consequently a negative $y$ range. We have chosen to present our results with a positive $y$ notation} with the muon spectrometer~\cite{Aamodt:2011gj}. 
The detectors relevant for \upsi measurements in \pbpb collisions are described below.

The Silicon Pixel Detector, corresponding to the two innermost layers of the Inner Tracking System~\cite{Aamodt:2010aa}, is used for the primary vertex determination. 
The inner and outer layer cover the pseudo-rapidity ranges $|\eta|<2$ and $|\eta|<1.4$, respectively.
The V0 scintillator hodoscopes~\cite{Abbas:2013taa} provide the centrality estimate.
They are made of two arrays of scintillators placed in the pseudo-rapidity ranges $2.8<\eta<5.1$ and $-3.7<\eta<-1.7$. 
The logical AND of the signals from the two hodoscopes constitutes the Minimum Bias (MB) trigger. 
The MB trigger is fully efficient for the studied 0--90\% most central collisions.
The Zero Degree Calorimeters (ZDC) are installed at $\pm112.5$ m from the nominal interaction point along the beam line. 
Each of the two ZDCs is composed of two sampling calorimeters designed for detecting spectator protons, neutrons and nuclear fragments. 
The evaluation of the signal amplitudes of the ZDCs allows for the rejection of events corresponding to an electromagnetic interaction of the colliding Pb nuclei \cite{ALICE:2012aa}.

The muon spectrometer covers the pseudorapidity range $-4<\eta<-2.5$. It is composed of a front absorber, which filters muons upstream of the muon tracker, consisting of five tracking stations with two planes of cathode-pad chambers each, and of a dipole magnet providing a $3$ T$\cdot$m integrated magnetic field.
Downstream of the tracking system, a 1.2~m thick iron wall stops efficiently the punch-through hadrons. 
The muon trigger system is located downstream of the iron wall and consists of two stations, each one equipped with two planes of Resistive Plate Chambers (RPC), with an efficiency higher than 95\%~\cite{mtrBossu2012}.
The muon-trigger system is able to deliver single and dimuon triggers selecting muons with $\pt$ larger than a programmable threshold, via an algorithm based on the RPC spatial information~\cite{mtr2006}.
Throughout its entire length, a conical absorber shields the muon spectrometer against secondary particles produced by the interaction of primary particles in the beam pipe. 

The trigger condition used for data taking is a dimuon-Minimum Bias ($\mu\mu\text{-MB}$) trigger formed by the logical AND of the MB trigger and an unlike-sign dimuon trigger with a \pt threshold of 1~GeV/$c$ for each of the two muons. 

The centrality estimation is performed using a Glauber fit to the sum of the signal amplitudes of the V0 scintillators~\cite{Abelev:2013qoq,Adam:2015ptt,ALICE-PUBLIC-2018-011}. 
Centrality ranges are given as percentages of the total hadronic \pbpb cross section. 
In addition to the centrality, the Glauber model allows an estimate of the average number of participant nucleons $\langle N_{\rm part} \rangle$, of the average number of binary collisions $\langle N_{\rm coll} \rangle$ and of the nuclear overlap function $\langle T_{\rm AA} \rangle$, for each centrality interval \cite{dEnterria:2003xac}. 
In the present analysis, the data sample corresponds to an integrated luminosity $L_{\rm int}\approx 225$ $\mu$b$^{-1}$ in the centrality interval 0--90\% that has been divided into four centrality classes: 0--10\%, 10--30\%, 30--50\% and 50--90\%.

\section{Data analysis}\label{section:analysis}

The evaluation of $R_{\rm AA}$ is performed through the following expression:
\begin{equation} \label{eqn:raa}
R_{\rm{AA}} = \frac{N^{\Upsilon}}{ {\rm BR}_{\Upsilon \rightarrow \mu^{+} \mu^{-}} \cdot  (A\times\epsilon)_{\Upsilon \rightarrow \mu^{+} \mu^{-}} \cdot  N_{\mu\mu\text{-MB}} \cdot F_{\rm norm} \cdot \sigma_{\rm pp}^{\Upsilon} \cdot \langle T_{\rm AA} \rangle } ~~ ,
\end{equation}
where $N^{\Upsilon}$ is the number of detected resonance decays to muon pairs, while ${\rm BR}_{\Upsilon \rightarrow \mu^{+} \mu^{-}}=(2.48\pm0.05)$\% and $(1.93\pm0.17)$\% are the branching ratios for the dimuon decay of \upsis and $\Upsilon\text{(2S)}$, respectively~\cite{Patrignani:2016xqp}. 
The $(A\times\epsilon)_{\Upsilon \rightarrow \mu^{+} \mu^{-}}$ factor is the product of acceptance and detection efficiency for the $\Upsilon$ state under study. 
The normalization factor $N_{\mu\mu\text{-MB}} \cdot F_{\rm norm}$ is the product of the number of analyzed $\mu\mu\text{-MB}$ events and the inverse of the probability to obtain an unlike-sign dimuon trigger in a MB-triggered event~\cite{Adam:2016rdg}. 
A dataset of $1.5\cdot10^9$ minimum bias equivalent events, $N_{\mu\mu\text{-MB}} \cdot F_{\rm norm}$, has been used for bottomonium measurements.  
Finally, $\sigma_{\rm pp}^{\Upsilon}$ is the reference \pp cross section and $\langle T_{\rm AA} \rangle$ represents the nuclear overlap function~\cite{ALICE-PUBLIC-2018-011}.


\begin{figure}[!b]
\begin{center}
\includegraphics[width=0.8\linewidth]{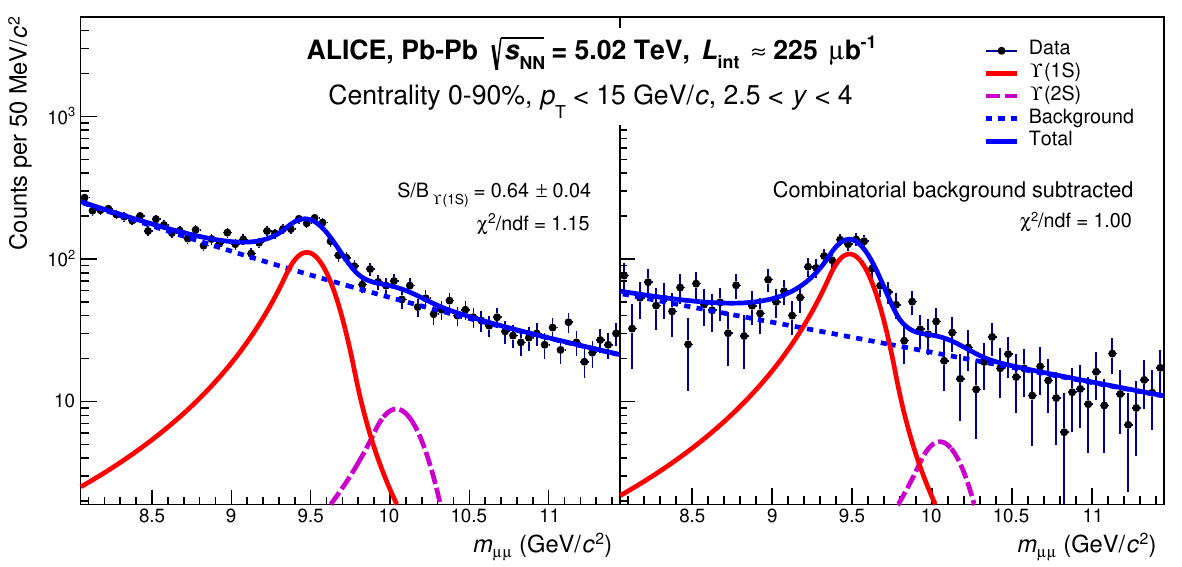}
\caption{Red and magenta solid lines correspond to \upsis and \upsiss signal functions, respectively. The contribution from \upsisss yield is compatible with zero. Dotted blue lines represent the background (left) and residual background (right), respectively. The sum of the various functions is also shown as a solid blue line.}
\label{fig:mass_dist}
\end{center}
\end{figure}

The signal yields are evaluated by performing fits to the $\mu^+\mu^-$ invariant mass distributions. 
In order to improve the purity of the dimuon sample a set of selection criteria~\cite{Abelev:2014nua} has been applied on the muon tracks, including the request of the matching between the tracks reconstructed in the trigger and tracking detectors of the muon spectrometer and a cut on the track transverse momentum ($p_{\rm T} > 2$ GeV/$c$).
The latter cut has a small effect ($\sim 2\%$) on the number of detected resonances. 
The raw \upsi yields are extracted using the sum of three extended Crystal Ball (CB) functions~\cite{ALICE-Quarkonia-signal-extraction}, one for each of \upsis, \upsiss and \upsisss.  
The extended CB function consists of a Gaussian core with  non-Gaussian tails on both sides to take into account the radiative contributions of the \upsi production and the absorber effects of muon energy loss in the low mass tail, whereas the high mass tail is attributed to the multiple Coulomb scattering in the front absorber and the momentum resolution of the tracking chambers.    
The background is fitted with the sum of two exponential functions (see left panel of Fig.~\ref{fig:mass_dist}).
Since the signal-to-background (S/B) ratio is low in the tail regions of the extended CB functions, the tail parameters are fixed to values obtained from the Monte Carlo (MC) simulation. 
The mass position and the width parameters of the \upsis are left free for the integrated spectrum ($i.e.$ centrality class 0--90\%, $p_{\rm T}<15$ GeV/$c$ and $2.5 < y < 4$).
Whereas for the signal extraction as a function of centrality, the mass position  and width ($160 \pm 15$ MeV/$c^2$) of the \upsis are fixed to the values obtained in the fit to the centrality-integrated (0--90\%) mass spectrum. 
Finally, for studies as a function of \pt and $y$, the mass position and the width obtained for the centrality-integrated mass spectrum are scaled according to their evolution observed in the MC.
Due to the poor S/B ratio for the higher-mass states, the values of the mass of the \upsiss and \upsisss are fixed to the PDG mass differences with respect to the \upsis, and the ratio of \upsiss ($\Upsilon\text{(3S)}$) to \upsis widths is fixed to values from the MC simulation, $i.e.$ 1.03 (1.06). 
In the fit shown in Fig.~\ref{fig:mass_dist} only signals corresponding to the $\Upsilon\text{(1S)}$ and $\Upsilon\text{(2S)}$ are visible, since the $\Upsilon\text{(3S)}$ contribution is compatible with zero events.
Alternatively, the combinatorial background is modeled with the event-mixing method. 
In this approach, an invariant mass dimuon spectrum is constructed by pairing muons from different events with similar multiplicities as described in~\cite{Adam:2016rdg}.
The combinatorial background is then subtracted from the raw dimuon spectrum (right panel of Fig.~\ref{fig:mass_dist}) and the resulting distribution is fitted with the sum of three extended CB and an exponential function to account for the residual background. 
Finally, the number of detected \upsi resonances, $N^{\Upsilon}$, is obtained as the average~\cite{ALICE-Quarkonia-signal-extraction} of the fitting methods described above (and also below in the discussion on signal systematics), leading to $N^{\Upsilon{\rm (1S)}}$ = $1126\pm53\text{(stat)}\pm47\text{(syst)}$ and $N^{\Upsilon{\rm (2S)}}$ = $77\pm33\text{(stat)}\pm17\text{(syst)}$.


The measured \upsi yields, $N^{\Upsilon}$, are corrected for the detector acceptance and efficiency using MC simulations.
Since the occupancy of the detector varies with the centrality of the collisions, the generated \upsi decays are embedded into real MB events to simulate the various particle-multiplicity scenarios as in data. 
The $p_{\rm T}$ and $y$ distributions of the generated \upsi are obtained from existing pp measurements ~\cite{Acosta:2001gv,LHCb:2012aa,Khachatryan:2010zg} using the interpolation procedure described in~\cite{Bossu:2011qe}. 
The EKS98 nuclear shadowing parameterization~\cite{Eskola:1998df} is used to include an estimate of CNM effects. 
Since available data favor a small or null polarization for $\Upsilon$(1S)~\cite{Abazov:2008aa,CDF:2011ag,Chatrchyan:2012woa,Aaij:2017egv}, an unpolarized production is assumed. 
The variations of the performance of the tracking and triggering systems throughout the data-taking period as well as the residual misalignment of the tracking chambers are taken into account in the simulation.
The $A\times\epsilon$ values, for the range $p_{\rm T} < 15$ GeV/$c$, $2.5 < y < 4$ and the 0--90\% centrality class are $0.263$ and $0.264$ for the \upsis and \upsiss, respectively, with a negligible statistical uncertainty. 
A decrease of 2\% is observed in  $A\times\epsilon$ for the 0--10\% central collisions with respect to the 50--90\% sample due to the higher occupancy in the most central events.  
The $A\times\epsilon$ is higher by 20\% in $3 < y < 3.5$  compared to the values at $2.5 < y < 3$ and $3.5 < y < 4$ mainly due to the geometric acceptance of the detector, whereas it has no variation as a function of \pt. 
The systematic uncertainty on $A\times\epsilon$ is discussed below.


The systematic uncertainty on the signal extraction is evaluated using various functions for modelling the background shape, as well as adopting two fitting ranges, $i.e.$ (7--14) GeV/$c^2$ and (7.5--14.5) GeV/$c^2$. 
The tail parameters of the signal functions have been varied using estimates provided by two MC particle transport models: GEANT4~\cite{Agostinelli:2002hh} and GEANT3~\cite{Brun:1082634}. 
In the centrality, \pt or $y$ differential studies, the mass position and width are also varied by amounts, which correspond to the uncertainties on the mass position and the width returned by the fit to the centrality-integrated invariant mass spectrum. 
The ratio of \upsiss ($\Upsilon\text{(3S)}$) to \upsis widths is varied from 1 (1) to 1.06 (1.12).
The values of $N^{\Upsilon}$ and their statistical uncertainties are obtained by taking the average of $N^{\Upsilon}$ and of the corresponding statistical uncertainties from the various fits. 
This procedure is applied to both fits of the raw and combinatorial-background subtracted spectra. 
The systematic uncertainties are estimated as the root mean square of the distribution of $N^{\Upsilon}$ obtained from the various fits.
The effect induced by the $p_{\rm T} > 2$ GeV/$c$ cut on single muons on the $A\times\epsilon$-corrected \upsi yields was estimated by varying that cut by $\pm$0.2 GeV/$c$ in the MC.
A $\pm$2\% maximum variation on $N^{\Upsilon}/(A\times\epsilon)$ was observed and included in the systematic uncertainties.

Various sources contribute to the systematic uncertainties of $A\times\epsilon$, such as the \pt and $y$ shapes of the input distributions for the MC simulations, the trigger efficiency, the track reconstruction efficiency and finally the matching efficiency between tracks in the muon tracking and triggering chambers. 
Various sets of simulations are produced with different \upsi input \pt and $y$ distributions, obtained from empirical parameterizations and/or extrapolations of available data sets at different energies. 
The maximum relative difference of $A\times\epsilon$ for the various shapes is taken as the systematic uncertainty due to the input MC. 
In order to calculate the systematic uncertainty on trigger efficiency, the trigger response function for single muons is evaluated using either MC or data.
The two response functions are then separately applied to simulations of an \upsi sample and the difference obtained for the \upsi reconstruction efficiency is taken as systematic uncertainty.
The systematic uncertainty on the tracking efficiency is obtained starting from an evaluation of the single muon tracking efficiency in MC and data. 
This evaluation is performed via a procedure, detailed in ~\cite{Adam:2016rdg}, based on the redundancy of the tracking-chamber information.
The dimuon tracking efficiency is then obtained by combining the single muon efficiencies and the systematic uncertainty is taken as the difference of the values obtained with the procedure based on MC and data.
The muon tracks for data analysis are chosen based on a selection on the $\chi^2$ of the matching between a track segment in the trigger system with a track in the tracking chambers. 
The matching systematics are obtained by varying the $\chi^2$ selection cut in data and MC and comparing the effects on the muon reconstruction efficiency~\cite{Adam:2016rdg}.

The systematic uncertainty on the centrality measurement is evaluated by varying the V0 signal amplitude by $\pm0.5\%$ corresponding to 90\% of the hadronic cross section in \pbpb collisions, used as anchor point to define the centrality classes.
The systematic uncertainty on the evaluation of $\sigma^{\Upsilon}_{\rm pp}$ is detailed in the next section. 
Finally, the systematic-uncertainty evaluation of $F_{\rm{norm}}$ and $\langle T_{\rm{AA}}\rangle$ are described in~\cite{Adam:2016rdg} and~\cite{Abelev:2013qoq}, respectively.
The different systematic-uncertainty sources on the $\raa$ calculation are summarized in Table~\ref{tab:syst}.
If the above mentioned systematic uncertainty is correlated as a function of centrality, $\pt$ or $y$, it is quoted as correlated (type I) systematic uncertainty, otherwise it is treated as uncorrelated (type II).  

\begin{table}[!t]
\centering
\resizebox{1\textwidth}{!}{
\begin{tabular}{l l l l l l}
\hline
\multirow{2}{*}{Sources} & \multicolumn{4}{c}{$\upsis$} & $\upsiss$ \\
\cline{2-5}
& Centrality & $y$ & $\pt$ & Integrated & Integrated \\
\hline
Signal extraction                               		& 4.3-6.1\%(II)                   	& 4.2-6.8\%(II)                  	& 5.2-8.7\%(II)                	& 4.1\%  	& 21.7\%\\
Muon \pt cut                               		& 0.3-2.4\%(II)                   	& 0.1-1.2\%(II)                  	& 0.1-2.4\%(II)                	& 0.7\%  	& 0.7\%\\
Input MC                                        		& 0.9\%(I)                       		& 0.6-2.6\%(II)                   & 1-1.4\%(II)                	& 0.9\%  	& 0.9\%\\
Tracker efficiency                             		& 3\%(I) and 0-1\%(II)      		& 1\%(I) and 3\%(II)		& 1\%(I) and 3\%(II)   	& 3\%  	& 3\%\\
Trigger efficiency                                       	& 3\%(I)                        		& 1.4-3.7\%(II)                   & 0.4-2.6\%(II)                 	& 3\%  	& 3\%\\
Matching efficiency                                     & 1\%(I)                        		& 1\%(II)                     	& 1\%(II)	                  	& 1\%  	& 1\%\\
Centrality                                      		& 0.2-2.4\%(II)                     	& -                      		& -	                   		& -  		& - \\
$F_{\rm{norm}}$                               		& 0.5\%(I)                        		& 0.5\%(I)                     	& 0.5\%(I)	                  	& 0.5\%  	& 0.5\%\\
$\langle T_{\rm{AA}}\rangle$                  	& 1.9-2.8\%(II)                      	& 2.4\%(I)                     	& 2.4\%(I)	                   	& 2.4\% 	& 2.4\%\\
${\rm BR}_{\Upsilon \rightarrow \mu^{+} \mu^{-}} \cdot \sigma^{\rm pp}_{\Upsilon}$                 	& 6.3\%(I)		                        	& 6.6-11.3\%(II)                 	& 5.5-11.5\%(II)		        	& 6.3\% 	& 7.5\%\\
\hline
\end{tabular}
}
\caption{\label{tab:syst}Summary of the systematic uncertainties for the $\raa$ calculation. Type I (II) refers to correlated (uncorrelated) systematic uncertainties.}
\end{table}

\section{Proton-proton reference cross sections}

The pp reference cross sections for \upsis and \upsiss production are computed by means of an interpolation procedure as described for \upsis in~\cite{LHCb:2014fxt}. 
The energy interpolation for the \upsi cross section, as a function of rapidity and for the \pt and $y$ integrated result, uses the measurements of \upsi production cross sections in \pp collisions at $\sqrt{s}=7$ and 8 TeV by ALICE \cite{Abelev:2014qha,Adam:2015rta} and at $\sqrt{s}=2.76, 7$ and 8 TeV by LHCb \cite{Aaij:2014nwa,Aaij:2015awa}. 
The interpolation is performed by using various empirical functions and, in addition, the shape of the energy dependence of the bottomonium cross sections calculated using two theoretical models, i.e. the Leading Order Colour Evaporation Model (LO-CEM) \cite{Cheung:2017osx}  and the Fixed Order Next-to-Leading Logs (FONLL) model \cite{Cacciari:1998it}.
The latter gives cross sections for open beauty, which is here used as a proxy to study the evolution of the bottomonium cross section~\cite{LHCb:2014fxt}. 
The energy interpolation for the \upsis cross section as a function of \pt is based on LHCb measurements only, since the  \pt coverage of the results of this analysis ($p_{\rm T} < 15$ GeV/$c$) is more extended than that of the corresponding ALICE pp data ($p_{\rm T} < 12$ GeV/$c$).
The result of the interpolation procedure gives ${\rm BR}_{\Upsilon\text{(1S)} \rightarrow \mu^{+} \mu^{-}} \cdot \sigma^{\rm\Upsilon({\rm 1S})}_{\rm pp}=1221\pm77\text{(syst)}$~pb and ${\rm BR}_{\Upsilon\text{(2S)} \rightarrow \mu^{+} \mu^{-}} \cdot \sigma^{\rm\Upsilon({\rm 2S})}_{\rm pp}=302\pm23\text{(syst)}$~pb assuming unpolarized quarkonia and integrating over the
ranges $2.5 < y < 4$ and $p_{\rm T} < 15$ GeV/$c$.
The uncertainties correspond to the quadratic sum of two terms. 
The first term dominates the total uncertainty on the interpolated value and reflects the statistical and systematic uncertainties on the data points used in the interpolation procedure. 
The second term is related to the spread among the interpolated cross sections obtained by using either the empirical functions or the energy dependence estimated from the theoretical models mentioned above. 
The numerical values obtained from the interpolation procedure are summarized in Table~\ref{tab:ref_cs_pp} for the various kinematic ranges used in the analysis.

\begin{table}[!t]
\centering
\begin{tabular}{c | c | c}
\hline
\pt (GeV/$c$) & $y$ & ${\rm BR}_{\Upsilon\text{(1S)} \rightarrow \mu^{+} \mu^{-}} \cdot \sigma_{\rm pp}^{\Upsilon{\rm (1S)}}$ (pb) \tabularnewline
\hline 
{[}0-2{]} & \multirow{4}{*}{{[}2.5-4{]}} & $226 \pm 26$\tabularnewline
{[}2-4{]} &  & $361 \pm 20$\tabularnewline
{[}4-6{]} &  & $288 \pm 24$\tabularnewline
{[}6-15{]} &  & $311 \pm 23$\tabularnewline
\hline 
\multirow{3}{*}{{[}0-15{]}} & {[}2.5-3{]} & $ 506 \pm 57 $\tabularnewline
 & {[}3-3.5{]} & $ 415 \pm 28 $\tabularnewline
 & {[}3.5-4{]} & $ 288 \pm 24$\tabularnewline
\hline 
\end{tabular}
\caption{\label{tab:ref_cs_pp} The interpolated branching ratio times cross section of \upsis for the \pt and $y$ bins under study. The quoted uncertainties are systematic.}
\end{table}

\section{Results}\label{section:results}

The nuclear modification factors for  inclusive $\upsis$ and $\upsiss$ production in \pbpb collisions at $\snn=5.02$ TeV for the ranges $p_{\rm T} < 15$ GeV/$c$, $2.5 < y < 4$ and the 0--90\% centrality class are $R_{\rm{AA}}^{\Upsilon(\rm1S)}=0.37\pm0.02\text{(stat)}\pm0.03\text{(syst)}$ and $R_{\rm{AA}}^{\Upsilon(\rm2S)}=0.10\pm{0.04}\text{(stat)}\pm{0.02}\text{(syst)}$, respectively. 
The ratio $R_{\rm{AA}}^{\Upsilon(\rm2S)}/R_{\rm{AA}}^{\Upsilon(\rm1S)}$ is $0.28\pm0.12\text{(stat)}\pm0.06\text{(syst)}$. 
Since the decay kinematics of the two \upsi states is very similar, most of the systematic uncertainty sources entering the ratio cancel out except those on the signal extraction and on the pp cross section, which are the dominant contributions to the total systematic uncertainty. 
The measurements show a strong suppression for both bottomonium states with the more weakly bound state being significantly more suppressed.  
The ratio between the \upsis $\raa$ at $\snn=5.02$ TeV and 2.76 TeV~\cite{Abelev:2014nua} is $1.23\pm0.21\text{(stat)}\pm0.19\text{(syst)}$. 
The sources of systematic uncertainties entering the calculation of the ratio are considered uncorrelated, except for the \taa component, whose uncertainty cancels out. 
The ratio is compatible with unity within uncertainties.

\begin{figure}[!t]
\begin{center}
\includegraphics[width=0.65\linewidth]{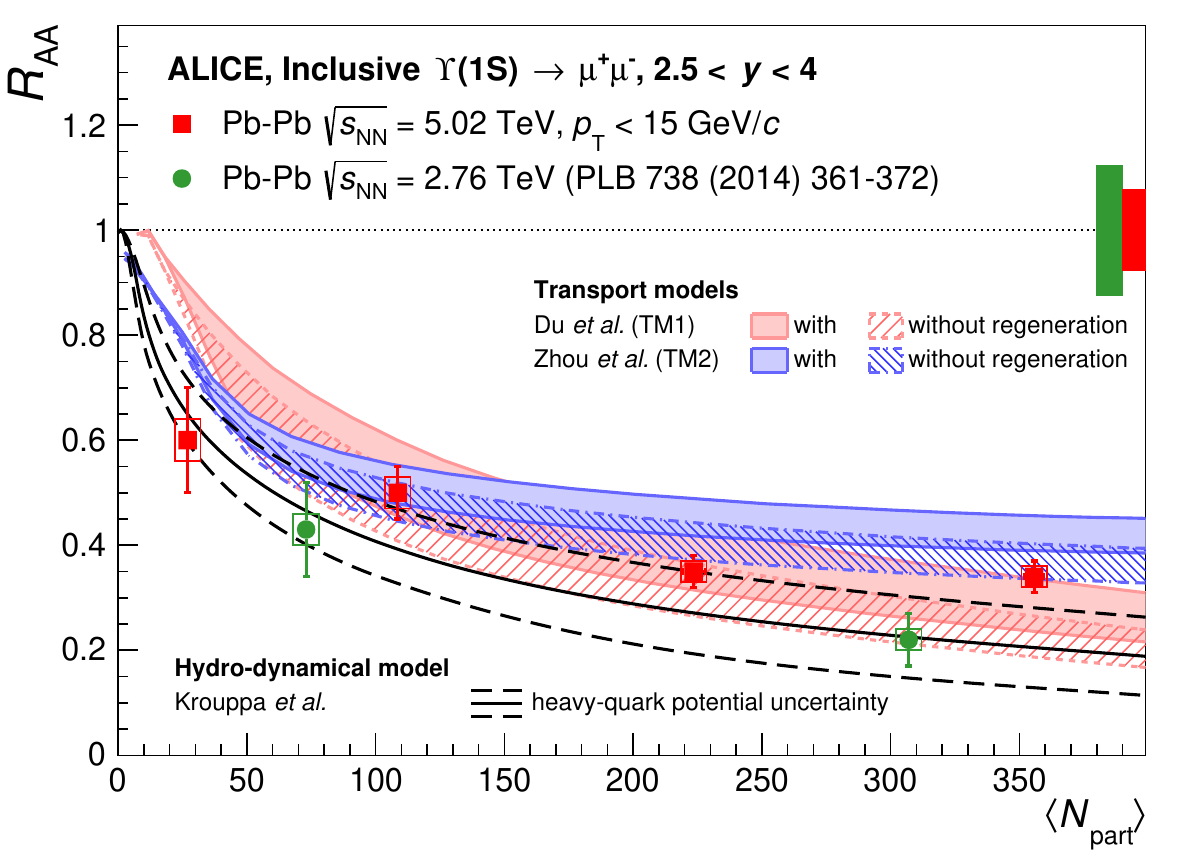} \\
\end{center}
\includegraphics[width=0.5\linewidth]{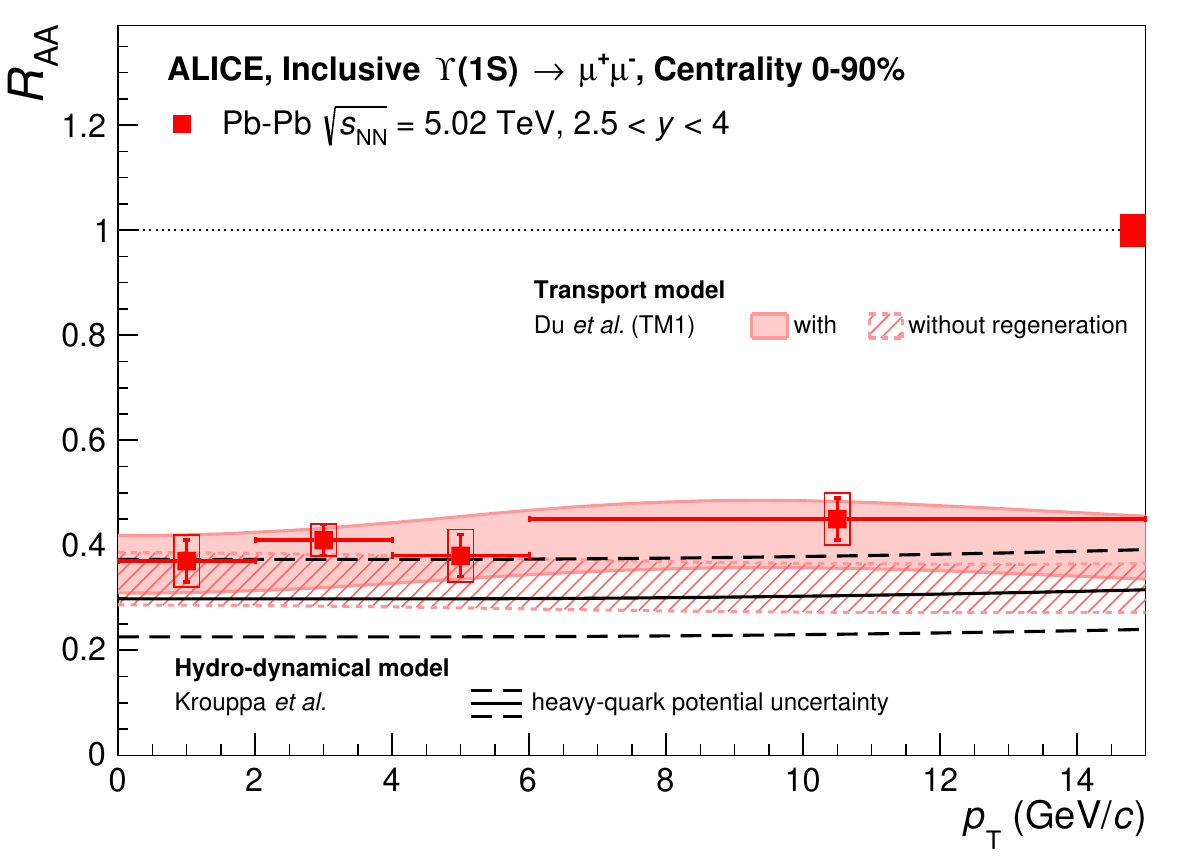}
\includegraphics[width=0.5\linewidth]{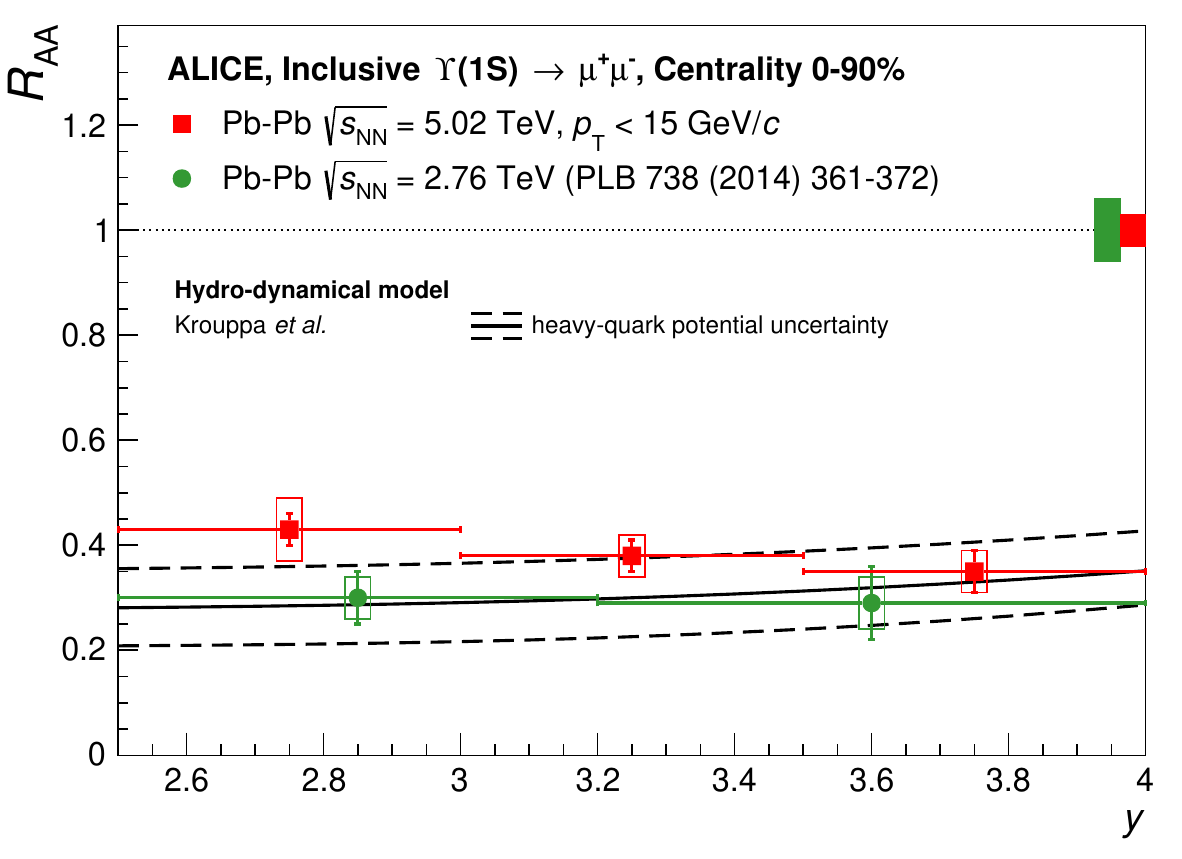}
\caption{Inclusive $\upsis$ $\raa$ as a function of centrality (top), $\pt$ (left) and $y$ (right) at forward rapidity at $\snn=5.02$ TeV. ALICE results at $\snn=2.76$ TeV as a function of centrality and $y$ are shown for comparison~\cite{Abelev:2014nua}. The vertical error bars and the boxes represent the statistical and uncorrelated systematic uncertainties, respectively. The relative correlated uncertainty is shown as boxes at unity. ALICE $\upsis$ $\raa$ measurements at $\snn=5.02$ TeV are compared to predictions from two transport models \cite{Du:2017qkv,Zhou:2014hwa} and one hydro-dynamical model \cite{Krouppa:2017jlg} as a function of centrality (top), $\pt$ (left) and $y$ (right). See text for details on the models.}
\label{fig:raa_models}
\end{figure}

The centrality, $p_{\rm T}$ and $y$ dependences of the $\upsis$ $\raa$ at forward rapidity at $\snn=5.02$ TeV are shown in Fig.\ref{fig:raa_models}. 
A decrease of $\raa$ with increasing centrality is observed down to $R_{\rm{AA}}^{\Upsilon(\rm1S)}=0.34\pm0.03\text{(stat)}\pm0.02\text{(syst)}$ for the 0--10\% most central collisions.
No significant $p_{\rm T}$-dependence is observed up to $p_{\rm T}=15$ GeV/$c$ within uncertainties.
The nuclear modification factor shows no significant dependence on rapidity. 
The $\upsis$ $\raa$ as a function of centrality and rapidity measured by ALICE at $\snn=2.76$ TeV~\cite{Abelev:2014nua} are also shown in Fig.\ref{fig:raa_models}. Similar trends can be observed at both collision energies. 

The inclusive $\upsis$ $\raa$ measurements are compared in Fig.~\ref{fig:raa_models} to several calculations: two transport models (TM) \cite{Du:2017qkv,Zhou:2014hwa} and one hydro-dynamical model \cite{Krouppa:2017jlg}. To describe the quarkonium motion in the medium, both transport codes use a rate-equation approach which accounts for both suppression and (re)generation mechanisms in the QGP. In the TM1 model \cite{Du:2017qkv} the evolution of the thermal medium is based on a thermal-fireball expansion while the TM2 model \cite{Zhou:2014hwa} uses a 2+1 dimensional version of the ideal hydrodynamic equations.
The two models use different rate equations and both models include a feed-down contribution from higher-mass bottomonia to the \upsis.
In TM2, two sets of feed-down fractions are assumed. 
Finally, the \upsis production cross section in pp collisions at $\sqrt{s}=5.02$ TeV in the rapidity range $2.5 < y < 4$ is taken as $\text{d}\sigma^{\rm\Upsilon({\rm 1S})}_{\rm pp}/\text{d}y=28.8$ nb in TM1 and $\text{d}\sigma^{\rm\Upsilon({\rm 1S})}_{\rm pp}/\text{d}y=30$ nb in TM2.
Those values deviate by about $2 \sigma$ (TM1) and $1.4 \sigma$ (TM2) from the result obtained using the pp interpolation method reported in the previous section.
TM1 predictions are shown as bands accounting for shadowing effects as calculated in ~\cite{Eskola:2009uj}. 
The upper limit shown in Fig.~\ref{fig:raa_models} corresponds to the extreme case of the absence of shadowing while the lower limit reflects a reduction of $30\%$ due to shadowing. 
The TM1 model implements the feed-down fractions reported in \cite{Andronic:2015wma}.
In the TM2 model, the shadowing parameterization is based on EKS98 \cite{Eskola:1998df} and the band edges correspond to two different sets of feed-down fractions (27\% from $\chi_{\rm b}$; 11\% from $\Upsilon({\rm 2S}+{\rm 3S})$ and 37\% from $\chi_{\rm b}$; 12\% from $\Upsilon({\rm 2S}+{\rm 3S})$) adopted by the authors. 
In the third model \cite{Krouppa:2017jlg}, a thermal suppression of the bottomonium states is calculated using a complex-valued  heavy-quark potential parametrized by means of lattice QCD and embedded in a medium evolving according to 3+1d anisotropic hydrodynamics. 
In this recent study, the \raa shows no sensitivity to the plasma shear viscosity-to-entropy density ratio ($4\pi\eta/s$) parameter of the hydro evolution, which is therefore set to $4\pi\eta/s$ = 2 consistent with particle spectra fits.
The band of the model quantifies the heavy-quark potential uncertainty, which has been estimated by including a $\pm 15\%$ variation of the Debye mass of the QCD medium that is tuned by a fit to the real-part of the lattice in-medium heavy-quark potential.
Furthermore, the predictions shown are referring to the initial momentum-space anisotropy parameter $\xi_0=0$, which corresponds to a perfectly isotropic QGP at the starting point of the hydro-dynamical evolution at $\tau_0=0.3$ fm/$c$.
Finally, this model accounts for feed-down contributions but it includes neither a (re)generation mechanism nor CNM effects. 
The centrality dependence of the $\upsis$ $\raa$ is fairly reproduced by the model calculations in the top panel of Fig.~\ref{fig:raa_models}.
The data are best described by TM1 when (re)generation is included and by TM2 when (re)generation is not taken into account. 
The hydro-dynamical model describes the trend of the data, the fact that the data lie on the upper edge of the uncertainty band for $\npart > 70$ could indicate a smaller Debye mass and thus a stronger heavy-quark potential.
The data  as a function of \pt (bottom left panel of Fig.~\ref{fig:raa_models}) can be described with or without the (re)generation scenario of the TM1 model while showing agreement with the hydro-dynamical model for the upper edge of the uncertainty band.
Finally, the $y$-dependence of the $\upsis$ $\raa$ is described, within uncertainties, by the hydro-dynamical model in the bottom right panel of Fig.~\ref{fig:raa_models} despite the possibly different trend between data and calculations. 

The low $\upsis$ $\raa$ reported in this Letter raises the important question whether direct \upsis are suppressed at LHC energies or only the feed-down contribution from higher mass states.  
However, the large uncertainties of the current measurements  of CNM effects \cite{Abelev:2014oea,Aaij:2014mza,Aaboud:2017cif} prevent a firm conclusion. 

\section{Summary}

The nuclear modification factors of inclusive \upsis and \upsiss production at forward rapidity ($2.5<y<4$) and for $p_{\rm T}<15$ GeV/$c$ in \pbpb collisions at $\sqrt{s_{_{\rm NN}}}=5.02$~TeV have been measured using the ALICE detector.
The observed \upsis suppression increases with the centrality of the collision and no significant variation is observed as a function of transverse momentum or rapidity. 
A larger suppression of the \upsiss bound state  compared to the ground state is also reported. 
Transport and dynamical model calculations reproduce qualitatively the centrality and kinematic dependence of the \upsis nuclear modification factor.

%
%

\newenvironment{acknowledgement}{\relax}{\relax}
\begin{acknowledgement}
\section*{Acknowledgements}

The ALICE Collaboration would like to thank all its engineers and technicians for their invaluable contributions to the construction of the experiment and the CERN accelerator teams for the outstanding performance of the LHC complex.
The ALICE Collaboration gratefully acknowledges the resources and support provided by all Grid centres and the Worldwide LHC Computing Grid (WLCG) collaboration.
The ALICE Collaboration acknowledges the following funding agencies for their support in building and running the ALICE detector:
A. I. Alikhanyan National Science Laboratory (Yerevan Physics Institute) Foundation (ANSL), State Committee of Science and World Federation of Scientists (WFS), Armenia;
Austrian Academy of Sciences and Nationalstiftung f\"{u}r Forschung, Technologie und Entwicklung, Austria;
Ministry of Communications and High Technologies, National Nuclear Research Center, Azerbaijan;
Conselho Nacional de Desenvolvimento Cient\'{\i}fico e Tecnol\'{o}gico (CNPq), Universidade Federal do Rio Grande do Sul (UFRGS), Financiadora de Estudos e Projetos (Finep) and Funda\c{c}\~{a}o de Amparo \`{a} Pesquisa do Estado de S\~{a}o Paulo (FAPESP), Brazil;
Ministry of Science \& Technology of China (MSTC), National Natural Science Foundation of China (NSFC) and Ministry of Education of China (MOEC) , China;
Ministry of Science and Education, Croatia;
Ministry of Education, Youth and Sports of the Czech Republic, Czech Republic;
The Danish Council for Independent Research | Natural Sciences, the Carlsberg Foundation and Danish National Research Foundation (DNRF), Denmark;
Helsinki Institute of Physics (HIP), Finland;
Commissariat \`{a} l'Energie Atomique (CEA) and Institut National de Physique Nucl\'{e}aire et de Physique des Particules (IN2P3) and Centre National de la Recherche Scientifique (CNRS), France;
Bundesministerium f\"{u}r Bildung, Wissenschaft, Forschung und Technologie (BMBF) and GSI Helmholtzzentrum f\"{u}r Schwerionenforschung GmbH, Germany;
General Secretariat for Research and Technology, Ministry of Education, Research and Religions, Greece;
National Research, Development and Innovation Office, Hungary;
Department of Atomic Energy Government of India (DAE), Department of Science and Technology, Government of India (DST), University Grants Commission, Government of India (UGC) and Council of Scientific and Industrial Research (CSIR), India;
Indonesian Institute of Science, Indonesia;
Centro Fermi - Museo Storico della Fisica e Centro Studi e Ricerche Enrico Fermi and Istituto Nazionale di Fisica Nucleare (INFN), Italy;
Institute for Innovative Science and Technology , Nagasaki Institute of Applied Science (IIST), Japan Society for the Promotion of Science (JSPS) KAKENHI and Japanese Ministry of Education, Culture, Sports, Science and Technology (MEXT), Japan;
Consejo Nacional de Ciencia (CONACYT) y Tecnolog\'{i}a, through Fondo de Cooperaci\'{o}n Internacional en Ciencia y Tecnolog\'{i}a (FONCICYT) and Direcci\'{o}n General de Asuntos del Personal Academico (DGAPA), Mexico;
Nederlandse Organisatie voor Wetenschappelijk Onderzoek (NWO), Netherlands;
The Research Council of Norway, Norway;
Commission on Science and Technology for Sustainable Development in the South (COMSATS), Pakistan;
Pontificia Universidad Cat\'{o}lica del Per\'{u}, Peru;
Ministry of Science and Higher Education and National Science Centre, Poland;
Korea Institute of Science and Technology Information and National Research Foundation of Korea (NRF), Republic of Korea;
Ministry of Education and Scientific Research, Institute of Atomic Physics and Romanian National Agency for Science, Technology and Innovation, Romania;
Joint Institute for Nuclear Research (JINR), Ministry of Education and Science of the Russian Federation and National Research Centre Kurchatov Institute, Russia;
Ministry of Education, Science, Research and Sport of the Slovak Republic, Slovakia;
National Research Foundation of South Africa, South Africa;
Centro de Aplicaciones Tecnol\'{o}gicas y Desarrollo Nuclear (CEADEN), Cubaenerg\'{\i}a, Cuba and Centro de Investigaciones Energ\'{e}ticas, Medioambientales y Tecnol\'{o}gicas (CIEMAT), Spain;
Swedish Research Council (VR) and Knut \& Alice Wallenberg Foundation (KAW), Sweden;
European Organization for Nuclear Research, Switzerland;
National Science and Technology Development Agency (NSDTA), Suranaree University of Technology (SUT) and Office of the Higher Education Commission under NRU project of Thailand, Thailand;
Turkish Atomic Energy Agency (TAEK), Turkey;
National Academy of  Sciences of Ukraine, Ukraine;
Science and Technology Facilities Council (STFC), United Kingdom;
National Science Foundation of the United States of America (NSF) and United States Department of Energy, Office of Nuclear Physics (DOE NP), United States of America.
\end{acknowledgement}

\bibliographystyle{utphys}   

\bibliography{draft}


\newpage
\appendix
\section{The ALICE Collaboration}
\label{app:collab}

\begingroup
\small
\begin{flushleft}
S.~Acharya\Irefn{org139}\And 
F.T.-.~Acosta\Irefn{org20}\And 
D.~Adamov\'{a}\Irefn{org93}\And 
J.~Adolfsson\Irefn{org80}\And 
M.M.~Aggarwal\Irefn{org98}\And 
G.~Aglieri Rinella\Irefn{org34}\And 
M.~Agnello\Irefn{org31}\And 
N.~Agrawal\Irefn{org48}\And 
Z.~Ahammed\Irefn{org139}\And 
S.U.~Ahn\Irefn{org76}\And 
S.~Aiola\Irefn{org144}
% ...
\And
\renewcommand\labelenumi{\textsuperscript{\theenumi}~}

\section*{Affiliation notes}
\renewcommand\theenumi{\roman{enumi}}
\begin{Authlist}
\item \Adef{org*}Deceased
\item \Adef{orgI}Dipartimento DET del Politecnico di Torino, Turin, Italy
\item \Adef{orgII}M.V. Lomonosov Moscow State University, D.V. Skobeltsyn Institute of Nuclear, Physics, Moscow, Russia
\item \Adef{orgIII}Department of Applied Physics, Aligarh Muslim University, Aligarh, India
\item \Adef{orgIV}Institute of Theoretical Physics, University of Wroclaw, Poland
\end{Authlist}

\section*{Collaboration Institutes}
\renewcommand\theenumi{\arabic{enumi}~}
\begin{Authlist}
\item \Idef{org1}A.I. Alikhanyan National Science Laboratory (Yerevan Physics Institute) Foundation, Yerevan, Armenia
\item \Idef{org2}Bogolyubov Institute for Theoretical Physics, National Academy of Sciences of Ukraine, Kiev, Ukraine
\item \Idef{org3}Bose Institute, Department of Physics  and Centre for Astroparticle Physics and Space Science (CAPSS), Kolkata, India
\item \Idef{org4}Budker Institute for Nuclear Physics, Novosibirsk, Russia
\item \Idef{org5}California Polytechnic State University, San Luis Obispo, California, United States
\item \Idef{org6}Central China Normal University, Wuhan, China
\item \Idef{org7}Centre de Calcul de l'IN2P3, Villeurbanne, Lyon, France
\item \Idef{org8}Centro de Aplicaciones Tecnol\'{o}gicas y Desarrollo Nuclear (CEADEN), Havana, Cuba
\item \Idef{org9}Centro de Investigaci\'{o}n y de Estudios Avanzados (CINVESTAV), Mexico City and M\'{e}rida, Mexico
\item \Idef{org10}Centro Fermi - Museo Storico della Fisica e Centro Studi e Ricerche ``Enrico Fermi', Rome, Italy
\item \Idef{org11}Chicago State University, Chicago, Illinois, United States
\item \Idef{org12}China Institute of Atomic Energy, Beijing, China
\item \Idef{org13}Chonbuk National University, Jeonju, Republic of Korea
\item \Idef{org14}Comenius University Bratislava, Faculty of Mathematics, Physics and Informatics, Bratislava, Slovakia
\item \Idef{org15}COMSATS Institute of Information Technology (CIIT), Islamabad, Pakistan
\item \Idef{org16}Creighton University, Omaha, Nebraska, United States
\item \Idef{org17}Department of Physics, Aligarh Muslim University, Aligarh, India
\item \Idef{org18}Department of Physics, Pusan National University, Pusan, Republic of Korea
\item \Idef{org19}Department of Physics, Sejong University, Seoul, Republic of Korea
\item \Idef{org20}Department of Physics, University of California, Berkeley, California, United States
\item \Idef{org21}Department of Physics, University of Oslo, Oslo, Norway
\item \Idef{org22}Department of Physics and Technology, University of Bergen, Bergen, Norway
\item \Idef{org23}Dipartimento di Fisica dell'Universit\`{a} 'La Sapienza' and Sezione INFN, Rome, Italy
\item \Idef{org24}Dipartimento di Fisica dell'Universit\`{a} and Sezione INFN, Cagliari, Italy
\item \Idef{org25}Dipartimento di Fisica dell'Universit\`{a} and Sezione INFN, Trieste, Italy
\item \Idef{org26}Dipartimento di Fisica dell'Universit\`{a} and Sezione INFN, Turin, Italy
\item \Idef{org27}Dipartimento di Fisica e Astronomia dell'Universit\`{a} and Sezione INFN, Bologna, Italy
\item \Idef{org28}Dipartimento di Fisica e Astronomia dell'Universit\`{a} and Sezione INFN, Catania, Italy
\item \Idef{org29}Dipartimento di Fisica e Astronomia dell'Universit\`{a} and Sezione INFN, Padova, Italy
\item \Idef{org30}Dipartimento di Fisica `E.R.~Caianiello' dell'Universit\`{a} and Gruppo Collegato INFN, Salerno, Italy
\item \Idef{org31}Dipartimento DISAT del Politecnico and Sezione INFN, Turin, Italy
\item \Idef{org32}Dipartimento di Scienze e Innovazione Tecnologica dell'Universit\`{a} del Piemonte Orientale and INFN Sezione di Torino, Alessandria, Italy
\item \Idef{org33}Dipartimento Interateneo di Fisica `M.~Merlin' and Sezione INFN, Bari, Italy
\item \Idef{org34}European Organization for Nuclear Research (CERN), Geneva, Switzerland
\item \Idef{org35}Faculty of Electrical Engineering, Mechanical Engineering and Naval Architecture, University of Split, Split, Croatia
\item \Idef{org36}Faculty of Engineering and Science, Western Norway University of Applied Sciences, Bergen, Norway
\item \Idef{org37}Faculty of Nuclear Sciences and Physical Engineering, Czech Technical University in Prague, Prague, Czech Republic
\item \Idef{org38}Faculty of Science, P.J.~\v{S}af\'{a}rik University, Ko\v{s}ice, Slovakia
\item \Idef{org39}Frankfurt Institute for Advanced Studies, Johann Wolfgang Goethe-Universit\"{a}t Frankfurt, Frankfurt, Germany
\item \Idef{org40}Gangneung-Wonju National University, Gangneung, Republic of Korea
\item \Idef{org41}Gauhati University, Department of Physics, Guwahati, India
\item \Idef{org42}Helmholtz-Institut f\"{u}r Strahlen- und Kernphysik, Rheinische Friedrich-Wilhelms-Universit\"{a}t Bonn, Bonn, Germany
\item \Idef{org43}Helsinki Institute of Physics (HIP), Helsinki, Finland
\item \Idef{org44}High Energy Physics Group,  Universidad Aut\'{o}noma de Puebla, Puebla, Mexico
\item \Idef{org45}Hiroshima University, Hiroshima, Japan
\item \Idef{org46}Hochschule Worms, Zentrum  f\"{u}r Technologietransfer und Telekommunikation (ZTT), Worms, Germany
\item \Idef{org47}Horia Hulubei National Institute of Physics and Nuclear Engineering, Bucharest, Romania
\item \Idef{org48}Indian Institute of Technology Bombay (IIT), Mumbai, India
\item \Idef{org49}Indian Institute of Technology Indore, Indore, India
\item \Idef{org50}Indonesian Institute of Sciences, Jakarta, Indonesia
\item \Idef{org51}INFN, Laboratori Nazionali di Frascati, Frascati, Italy
\item \Idef{org52}INFN, Sezione di Bari, Bari, Italy
\item \Idef{org53}INFN, Sezione di Bologna, Bologna, Italy
\item \Idef{org54}INFN, Sezione di Cagliari, Cagliari, Italy
\item \Idef{org55}INFN, Sezione di Catania, Catania, Italy
\item \Idef{org56}INFN, Sezione di Padova, Padova, Italy
\item \Idef{org57}INFN, Sezione di Roma, Rome, Italy
\item \Idef{org58}INFN, Sezione di Torino, Turin, Italy
\item \Idef{org59}INFN, Sezione di Trieste, Trieste, Italy
\item \Idef{org60}Inha University, Incheon, Republic of Korea
\item \Idef{org61}Institut de Physique Nucl\'{e}aire d'Orsay (IPNO), Institut National de Physique Nucl\'{e}aire et de Physique des Particules (IN2P3/CNRS), Universit\'{e} de Paris-Sud, Universit\'{e} Paris-Saclay, Orsay, France
\item \Idef{org62}Institute for Nuclear Research, Academy of Sciences, Moscow, Russia
\item \Idef{org63}Institute for Subatomic Physics, Utrecht University/Nikhef, Utrecht, Netherlands
\item \Idef{org64}Institute for Theoretical and Experimental Physics, Moscow, Russia
\item \Idef{org65}Institute of Experimental Physics, Slovak Academy of Sciences, Ko\v{s}ice, Slovakia
\item \Idef{org66}Institute of Physics, Homi Bhabha National Institute, Bhubaneswar, India
\item \Idef{org67}Institute of Physics of the Czech Academy of Sciences, Prague, Czech Republic
\item \Idef{org68}Institute of Space Science (ISS), Bucharest, Romania
\item \Idef{org69}Institut f\"{u}r Kernphysik, Johann Wolfgang Goethe-Universit\"{a}t Frankfurt, Frankfurt, Germany
\item \Idef{org70}Instituto de Ciencias Nucleares, Universidad Nacional Aut\'{o}noma de M\'{e}xico, Mexico City, Mexico
\item \Idef{org71}Instituto de F\'{i}sica, Universidade Federal do Rio Grande do Sul (UFRGS), Porto Alegre, Brazil
\item \Idef{org72}Instituto de F\'{\i}sica, Universidad Nacional Aut\'{o}noma de M\'{e}xico, Mexico City, Mexico
\item \Idef{org73}iThemba LABS, National Research Foundation, Somerset West, South Africa
\item \Idef{org74}Johann-Wolfgang-Goethe Universit\"{a}t Frankfurt Institut f\"{u}r Informatik, Fachbereich Informatik und Mathematik, Frankfurt, Germany
\item \Idef{org75}Joint Institute for Nuclear Research (JINR), Dubna, Russia
\item \Idef{org76}Korea Institute of Science and Technology Information, Daejeon, Republic of Korea
\item \Idef{org77}KTO Karatay University, Konya, Turkey
\item \Idef{org78}Laboratoire de Physique Subatomique et de Cosmologie, Universit\'{e} Grenoble-Alpes, CNRS-IN2P3, Grenoble, France
\item \Idef{org79}Lawrence Berkeley National Laboratory, Berkeley, California, United States
\item \Idef{org80}Lund University Department of Physics, Division of Particle Physics, Lund, Sweden
\item \Idef{org81}Nagasaki Institute of Applied Science, Nagasaki, Japan
\item \Idef{org82}Nara Women{'}s University (NWU), Nara, Japan
\item \Idef{org83}National and Kapodistrian University of Athens, School of Science, Department of Physics , Athens, Greece
\item \Idef{org84}National Centre for Nuclear Research, Warsaw, Poland
\item \Idef{org85}National Institute of Science Education and Research, Homi Bhabha National Institute, Jatni, India
\item \Idef{org86}National Nuclear Research Center, Baku, Azerbaijan
\item \Idef{org87}National Research Centre Kurchatov Institute, Moscow, Russia
\item \Idef{org88}Niels Bohr Institute, University of Copenhagen, Copenhagen, Denmark
\item \Idef{org89}Nikhef, National institute for subatomic physics, Amsterdam, Netherlands
\item \Idef{org90}NRC Kurchatov Institute IHEP, Protvino, Russia
\item \Idef{org91}NRNU Moscow Engineering Physics Institute, Moscow, Russia
\item \Idef{org92}Nuclear Physics Group, STFC Daresbury Laboratory, Daresbury, United Kingdom
\item \Idef{org93}Nuclear Physics Institute of the Czech Academy of Sciences, \v{R}e\v{z} u Prahy, Czech Republic
\item \Idef{org94}Oak Ridge National Laboratory, Oak Ridge, Tennessee, United States
\item \Idef{org95}Ohio State University, Columbus, Ohio, United States
\item \Idef{org96}Petersburg Nuclear Physics Institute, Gatchina, Russia
\item \Idef{org97}Physics department, Faculty of science, University of Zagreb, Zagreb, Croatia
\item \Idef{org98}Physics Department, Panjab University, Chandigarh, India
\item \Idef{org99}Physics Department, University of Jammu, Jammu, India
\item \Idef{org100}Physics Department, University of Rajasthan, Jaipur, India
\item \Idef{org101}Physikalisches Institut, Eberhard-Karls-Universit\"{a}t T\"{u}bingen, T\"{u}bingen, Germany
\item \Idef{org102}Physikalisches Institut, Ruprecht-Karls-Universit\"{a}t Heidelberg, Heidelberg, Germany
\item \Idef{org103}Physik Department, Technische Universit\"{a}t M\"{u}nchen, Munich, Germany
\item \Idef{org104}Research Division and ExtreMe Matter Institute EMMI, GSI Helmholtzzentrum f\"ur Schwerionenforschung GmbH, Darmstadt, Germany
\item \Idef{org105}Rudjer Bo\v{s}kovi\'{c} Institute, Zagreb, Croatia
\item \Idef{org106}Russian Federal Nuclear Center (VNIIEF), Sarov, Russia
\item \Idef{org107}Saha Institute of Nuclear Physics, Homi Bhabha National Institute, Kolkata, India
\item \Idef{org108}School of Physics and Astronomy, University of Birmingham, Birmingham, United Kingdom
\item \Idef{org109}Secci\'{o}n F\'{\i}sica, Departamento de Ciencias, Pontificia Universidad Cat\'{o}lica del Per\'{u}, Lima, Peru
\item \Idef{org110}Shanghai Institute of Applied Physics, Shanghai, China
\item \Idef{org111}St. Petersburg State University, St. Petersburg, Russia
\item \Idef{org112}Stefan Meyer Institut f\"{u}r Subatomare Physik (SMI), Vienna, Austria
\item \Idef{org113}SUBATECH, IMT Atlantique, Universit\'{e} de Nantes, CNRS-IN2P3, Nantes, France
\item \Idef{org114}Suranaree University of Technology, Nakhon Ratchasima, Thailand
\item \Idef{org115}Technical University of Ko\v{s}ice, Ko\v{s}ice, Slovakia
\item \Idef{org116}Technische Universit\"{a}t M\"{u}nchen, Excellence Cluster 'Universe', Munich, Germany
\item \Idef{org117}The Henryk Niewodniczanski Institute of Nuclear Physics, Polish Academy of Sciences, Cracow, Poland
\item \Idef{org118}The University of Texas at Austin, Austin, Texas, United States
\item \Idef{org119}Universidad Aut\'{o}noma de Sinaloa, Culiac\'{a}n, Mexico
\item \Idef{org120}Universidade de S\~{a}o Paulo (USP), S\~{a}o Paulo, Brazil
\item \Idef{org121}Universidade Estadual de Campinas (UNICAMP), Campinas, Brazil
\item \Idef{org122}Universidade Federal do ABC, Santo Andre, Brazil
\item \Idef{org123}University College of Southeast Norway, Tonsberg, Norway
\item \Idef{org124}University of Cape Town, Cape Town, South Africa
\item \Idef{org125}University of Houston, Houston, Texas, United States
\item \Idef{org126}University of Jyv\"{a}skyl\"{a}, Jyv\"{a}skyl\"{a}, Finland
\item \Idef{org127}University of Liverpool, Liverpool, United Kingdom
\item \Idef{org128}University of Tennessee, Knoxville, Tennessee, United States
\item \Idef{org129}University of the Witwatersrand, Johannesburg, South Africa
\item \Idef{org130}University of Tokyo, Tokyo, Japan
\item \Idef{org131}University of Tsukuba, Tsukuba, Japan
\item \Idef{org132}Universit\'{e} Clermont Auvergne, CNRS/IN2P3, LPC, Clermont-Ferrand, France
\item \Idef{org133}Universit\'{e} de Lyon, Universit\'{e} Lyon 1, CNRS/IN2P3, IPN-Lyon, Villeurbanne, Lyon, France
\item \Idef{org134}Universit\'{e} de Strasbourg, CNRS, IPHC UMR 7178, F-67000 Strasbourg, France, Strasbourg, France
\item \Idef{org135} Universit\'{e} Paris-Saclay Centre d¿\'Etudes de Saclay (CEA), IRFU, Department de Physique Nucl\'{e}aire (DPhN), Saclay, France
\item \Idef{org136}Universit\`{a} degli Studi di Foggia, Foggia, Italy
\item \Idef{org137}Universit\`{a} degli Studi di Pavia, Pavia, Italy
\item \Idef{org138}Universit\`{a} di Brescia, Brescia, Italy
\item \Idef{org139}Variable Energy Cyclotron Centre, Homi Bhabha National Institute, Kolkata, India
\item \Idef{org140}Warsaw University of Technology, Warsaw, Poland
\item \Idef{org141}Wayne State University, Detroit, Michigan, United States
\item \Idef{org142}Westf\"{a}lische Wilhelms-Universit\"{a}t M\"{u}nster, Institut f\"{u}r Kernphysik, M\"{u}nster, Germany
\item \Idef{org143}Wigner Research Centre for Physics, Hungarian Academy of Sciences, Budapest, Hungary
\item \Idef{org144}Yale University, New Haven, Connecticut, United States
\item \Idef{org145}Yonsei University, Seoul, Republic of Korea
\end{Authlist}
\endgroup
\end{document}